\documentclass[12pt]{iopart}

\usepackage{graphicx,color}
\usepackage{epsfig}
\usepackage{wrapfig}

\begin{document}

\title[Wibson, Rodrigues and Holanda .............................................................................]{Observation of Spin Current Hermiticity}

\author{Wibson W. G. Silva$^{1}$, A. R. Rodrigues$^{2}$ and  José Holanda$^{1,*}$}

\address{$^{1}$Programa de Pós-Graduação em Engenharia Física, Universidade Federal Rural de Pernambuco, 54518-430, Cabo de Santo Agostinho, Pernambuco, Brazil

$^{2}$Departamento de Física, Universidade Federal de Pernambuco, 50670-901, Recife, Pernambuco, Brazil}

\ead{$^{*}$joseholanda.silvajunior@ufrpe.br}

\vspace{10pt}
\begin{indented}
\item[]To arXiv
\end{indented}

\begin{abstract}
The spin Hall effect is one of the most relevant effects in spintronics and the key to conversion from charge current into spin current. We report here a phenomenon, which appears in response to the spin Hall effect and represents the anti-polarization of the spin current due to the coupling between interfacial magnetic anisotropies. Such an effect produces a Hermitian spin current. We realized experiments on permalloy (Py) and Cobalt (Co) bilayers to discuss this phenomenon.
\end{abstract}

In spintronics, the technique from broadband ferromagnetic resonance (FMR) microwave spectrometry has represented an experimental tool to study the dynamic properties of magnetic materials \cite{1, 2, 3}. Coupled with this conventional technique an electric current has been used to manipulate the spin current through the charge current in bilayers of magnetic and non-magnetic materials; such phenomenon is called spin Hall effect \cite{4, 5, 6, 7, 8, 9}. The result is the modulation of the damping of the magnetic material, opening new perspectives of applications in spintronics \cite{10, 11, 12}. The spin Hall effect represents the direct conversion from charge current into spin current due to spin-orbit interaction \cite{4, 5, 6, 7, 8, 9}. This phenomenon can also appear in the bilayers of magnetic materials. However, to date, there is no evidence of any anti-polarization observed from the spin current due to the coupling between magnetic interfacial anisotropies of two ferromagnetic materials with spin-orbit interaction \cite{13, 14}.

\begin{figure}[h]
\vspace{0.1mm} \hspace{0.1mm}
\begin{center}
\includegraphics[scale=0.42]{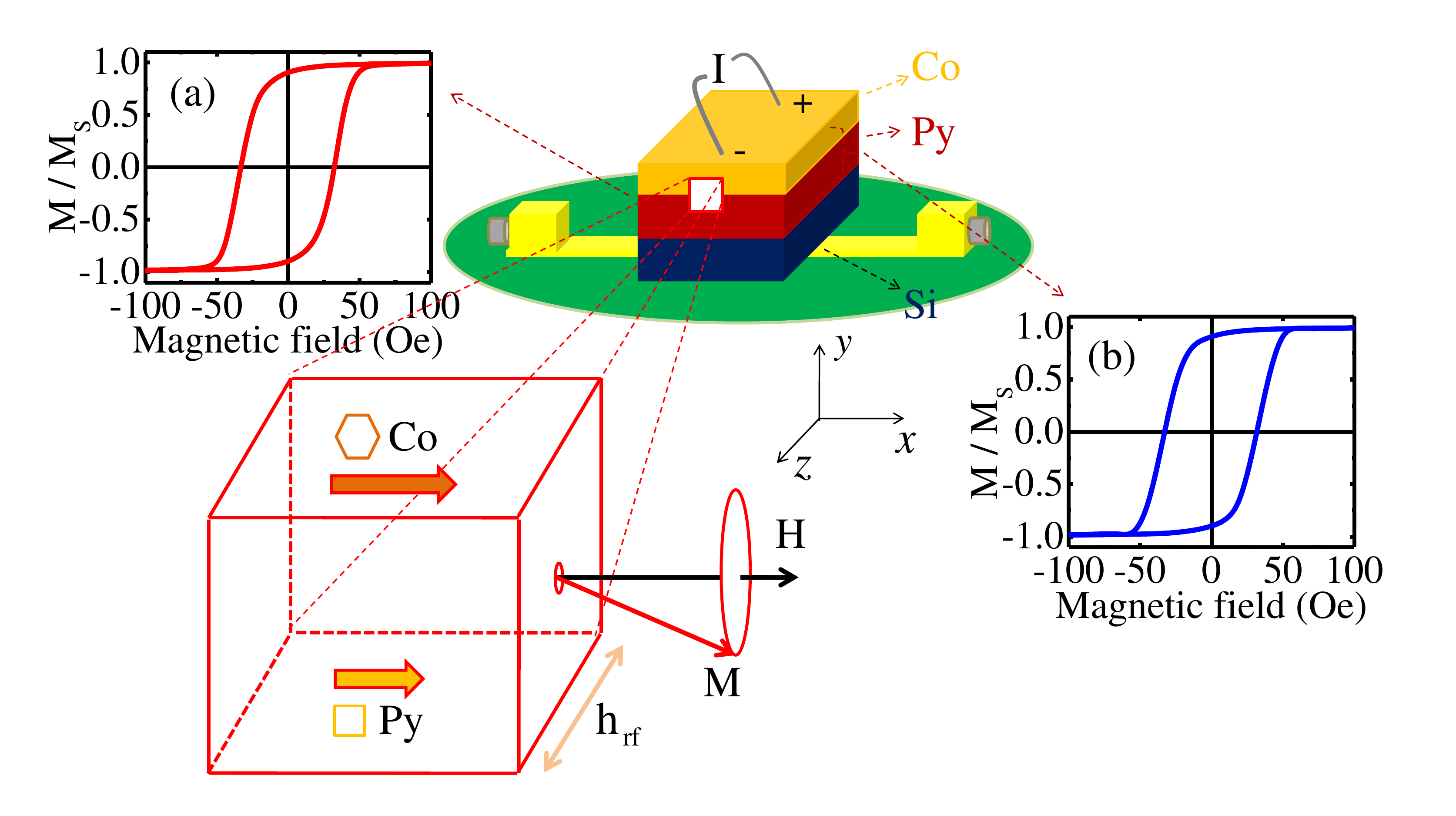}
\caption{\label{arttype}(Color on-line) Flip-chip ferromagnetic resonance setup and schematic of the dc current (\textit{I}) it is applied using Ag electrodes attached to the ends of the Co layer. The coupling between the two ferromagnetic layers is also schematically shown, where the intensity of the net magnetic moments of each material was evidenced. Typically, single Py (20 nm) films have hysteresis loops as shown. Hysteresis loops for: \textbf{(a)} Py (20 nm), and \textbf{(b)} Py(20 nm)/Co(10 nm).}
\label{puga}
\end{center}
\end{figure}

In this Letter, we related the observation of an anti-polarization from the spin current due to the coupling between local magnetic anisotropies, which arise from different spin configurations during the formation of magnetic domains \cite{15, 16, 17, 18, 19, 20, 21}. The samples used here represent 20 nm of Permalloy (Ni$_{80}$Fe$_{20}$) under 10 nm of cobalt (Co) and were grown on Si-substrate by magnetron sputtering technique at room and 300 °C temperatures for Py and Co, respectively. The Si-substrate was cut into a rectangular shape with dimensions of 4.0 $\times$ 2.0 mm$^{2}$. The 300 °C temperature of deposition of the Co was for a small diffusion on the Py film \cite{22}, which aimed to influence the distribution of local magnetic anisotropies at the interface. We performed magnetic resonance measurements without and with dc current in bilayers of permalloy (Ni$_{80}$Fe$_{20}$) and cobalt (Co), i. e., in samples of the type Py/Co, as shown in \textbf{Fig. \ref{puga}}. Typically, simple Py films show hysteresis loops as shown in \textbf{Fig. \ref{puga}  (a)}, which shows the appearance of interfacial magnetic anisotropies only after deposition from the Co film as evidenced in \textbf{Fig. \ref{puga} (b)}.

\begin{figure}[h]
	\vspace{0.5mm} \hspace{0.5mm}
	\begin{center}
		\includegraphics[scale=0.4]{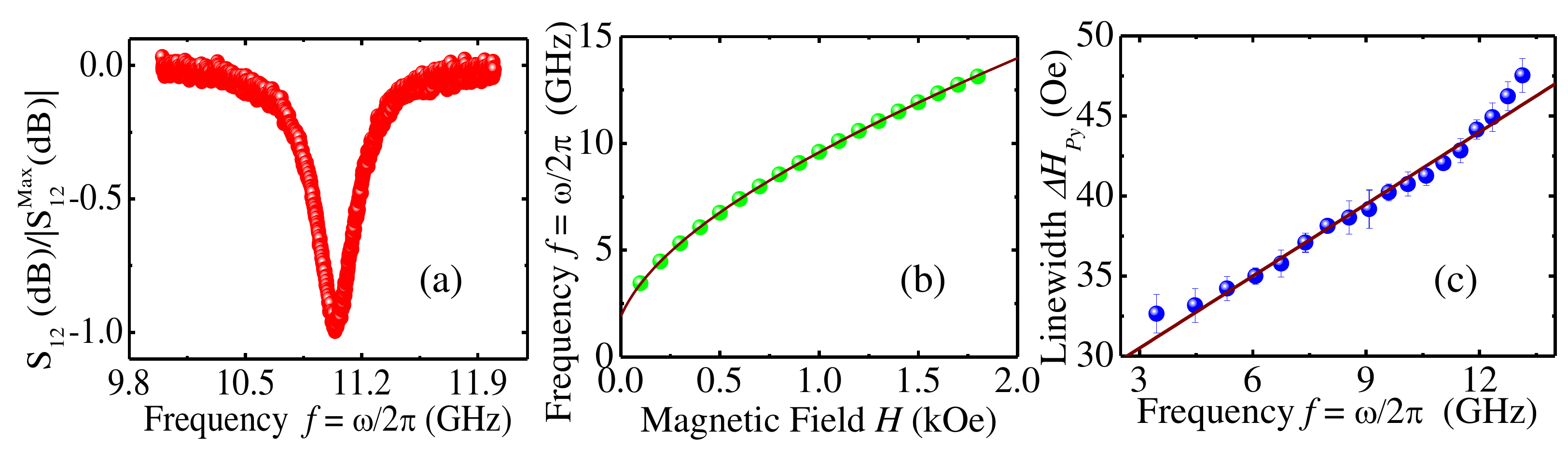}
		\caption{\label{arttype}(Color on-line) (a) Ferromagnetic resonance (FMR) signal obtained using the VNA for the magnetic field applied from \textit{H} = 1.3 kOe. (b) FMR frequency as a function of the magnetic field, where the solid line was obtained with the expression $f = \gamma \sqrt{(H + H_{A})(H + H_{A} + 4\pi M_{eff})}$, in which $\gamma$ = 2.8 GHz/kOe was used, where the anisotropic field of $H_A$ = 45 Oe and one effective saturation magnetization of $4\pi M_{eff}^{Py}$ = 10.2 kG. (c) The linewidth variation as a function of the FMR frequency, where the fit is shown was obtained with an intrinsic linewidth of $\Delta H_{0}$ = 26 Oe and damping of $\alpha_{Py} = 4.2 \times 10^{-3}$. All measurements are for single Py (20 nm) samples.}
		\label{gato}
	\end{center}
\end{figure}

We perform the measurements with transmission coefficient by sweeping the frequency at fixed fields. The frequency swept linewidths ($\Delta f_{VNA}$) we obtain via Lorentz function fitting. Detailed procedures, including the conversion from $\Delta f_{VNA}$ to magnetic field resonance linewidth $\Delta H$, can be found in Ref. \cite{23}. \textbf{Fig.\ref{gato} (a)} shows the FMR signal obtained with a vector network analyzer (VNA) for magnetic fields applied along the $\hat{x}$ direction without dc current for a single Py layer. This signal presents a linewidth of the order of $\Delta H$ = 42 Oe with one magnetic field of 1.3 kOe and one resonance frequency of 11 GHz. 

\begin{figure}[h]
	\vspace{0.5mm} \hspace{0.5mm}
	\begin{center}
		\includegraphics[scale=0.4]{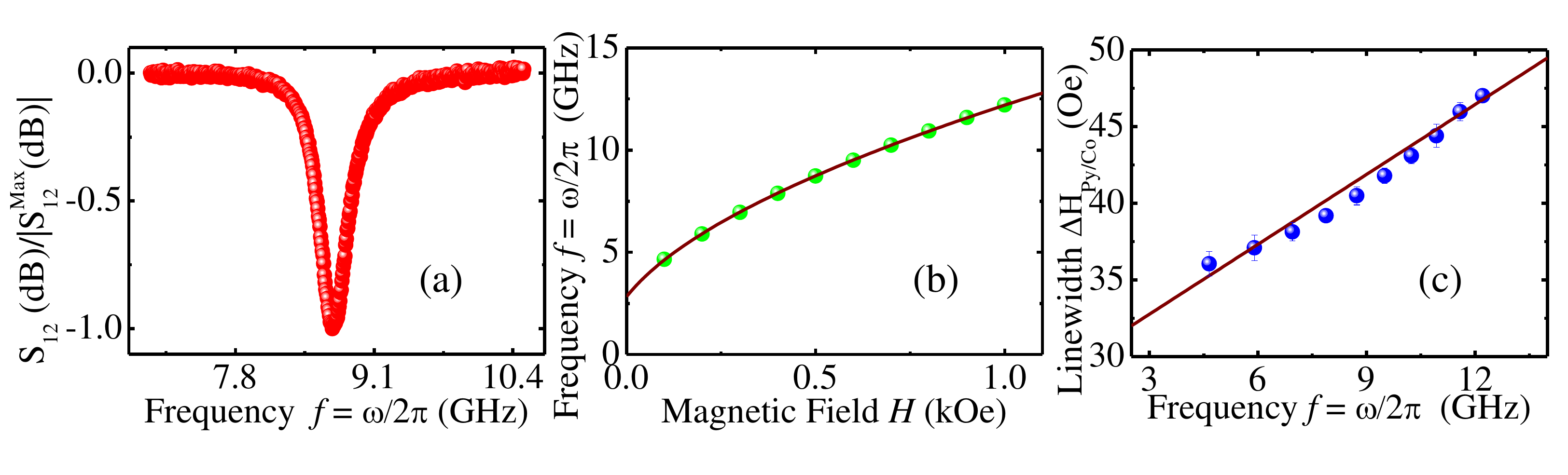}
		\caption{\label{arttype}(Color on-line) Measurements of the ferromagnetic resonance flip-chip for Py(20 nm)/Co(10 nm) samples. (a) Ferromagnetic resonance (FMR) signals were obtained using the VNA for the magnetic field applied from \textit{H} = 0.5 kOe. (b) FMR frequency as a function of the magnetic field, where the solid line was obtained with the expression $f = \gamma \sqrt{(H + H_{A})(H + H_{A} + 4\pi M_{eff})}$, in which $\gamma$ = 2.8 GHz/kOe was used, one anisotropic field of $H_A$ = 60.6 Oe and one effective saturation magnetization of $4\pi M_{eff}^{Py/Co}$ = 16.8 kG. (c) The linewidth variation as a function of the FMR frequency, where the fit was obtained with an intrinsic linewidth of $\Delta H_{0}^{Py/Co}$ = 28.2 Oe and damping of $\alpha_{Py/Co} = 4.3 \times 10^{-3}$.}
		\label{tatu}
	\end{center}
\end{figure}

In \textbf{Fig. \ref{gato} (b)} we present the resonance frequency as a function of the magnetic field, where the solid line was obtained with the expression $f = \gamma \sqrt{(H + H_{A})(H + H_{A} + 4\pi M_{eff})}$, in which $\gamma$ = 2.8 GHz/kOe was used, one anisotropic field of H$_A$ = 45 Oe and one effective saturation magnetization of $4\pi M_{eff}^{Py}$ = 10.2 kG. The linewidth as a function of the frequency is given by $\Delta H = \Delta H_{0} + (\alpha/\gamma)f$ \cite{24}, where $\alpha$ is the Gilbert damping and $\Delta H_{0}$ intrinsic linewidth
of the material. In \textbf{Fig. \ref{gato} (c)} we show the linewidth $\Delta H_{Py}$ as a function of frequency, where the fit was obtained with an intrinsic linewidth of $\Delta H_{0}^{Py}$ = 26 Oe and damping of $\alpha_{Py} = 4.2 \times 10^{-3}$. All parameters obtained for Py films are in agreement with the literature \cite{16}. The effects due to magnetic anisotropies are well known and can serve as spin current polarizers at interfaces with exchange bias \cite{25}, for example. Such a polarization phenomenon is so important that it is widely used in spin valves \cite{26, 27, 28}. At ferromagnetic interfaces, the polarization effect can present new characteristics. In \textbf{Fig. \ref{tatu} (a)} we show one from the FMR signals obtained with a vector network analyzer (VNA) for applied magnetic fields along the $\hat{x}$ direction without dc current for Py/Co samples.

In \textbf{Fig. \ref{tatu} (b)} we show the resonance frequency as a function of the magnetic field, in which one anisotropic field of $H_A$ = 60.6 Oe and one effective saturation magnetization of $4\pi M_{eff}^{Py/Co}$ = 16.8 kG was obtained. The increase in effective magnetization reveals the good coupling between the ferromagnetic layers \cite{22}. In \textbf{Fig. \ref{tatu} (c)} we plot the linewidth $\Delta H_{Py/Co}$ as a function of resonance frequency, whereby looking at the experimental data we get the intrinsic linewidth $\Delta H_{0}^{Py/Co}$ = 28.2 Oe and the damping $\alpha_{Py/Co} = 4.3 \times 10^{-3}$. We observed an increase in damping of Py/Co samples of the order of 2.3$\%$ compared to single Py samples. Such an increase is already well known for other materials and arises due to the formation of the interface between the layers \cite{15}. In the case presented here, the interface as also shown in ref. \cite{29}. 

\begin{figure}[h]
	\vspace{0.5mm} \hspace{0.5mm}
	\begin{center}
		\includegraphics[scale=0.4]{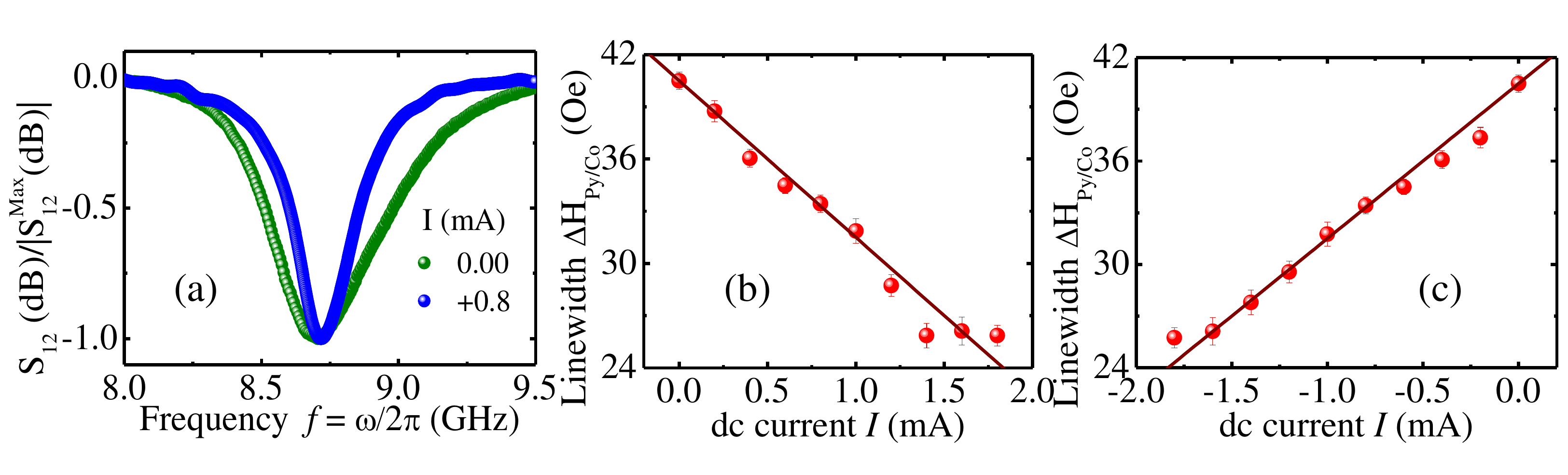}
		\caption{\label{arttype}(Color on-line) (a) Ferromagnetic resonance signals were obtained using the VNA for the magnetic field applied from H = 0.5 kOe and resonance frequency from 8.7 GHz without and with dc current from $I$ = + 0.8 mA. (b) and (c) Linewidth variation as a function of dc current positive and negative, respectively. The graph in (c) shows the effect of local anisotropies on spin current polarization. All measurements are for Py (20 nm)/Co(10 nm) samples applying dc current. The solid lines were obtained with the equations $\Delta H_{Py/Co} (I) = \Delta H_{0}^{Py/Co} \mp \beta I$ with $H_{0}^{Py/Co}$ = 40.5 Oe and $\beta$ = 8.7 Oe/A for (b) negative signal (positive dc current) and (c) positive signal (negative dc current), which show the linear modulation of the line width as a function of the dc current.}
		\label{cobra}
	\end{center}
\end{figure}

The analysis of linewidth variation as a function of dc current is shown in \textbf{Fig. \ref{cobra}} for one applied magnetic field  from \textit{H} = 0.5 kOe and one resonance frequency from 8.7 GHz. In \textbf{Fig. \ref{cobra} (a)} we show one from the FMR signals obtained with a vector network analyzer (VNA) for magnetic fields applied along the $\hat{x}$ direction without and with dc current from \textit{I} = +0.8 mA for Py/Co sample. The dependence of linewidth on dc current is shown in \textbf{Figs. \ref{cobra} (b)} and \textbf{(c)} for positive and negative dc currents, respectively. As for other bilayers, the linewidth decreases with the application of dc current to positive current \cite{4, 5, 6, 7, 8, 9} as shown in \textbf{Figs. \ref{cobra} (b)}.
However, it was expected that the linewidth would increase when the current polarization was reversed in the negative direction, which did not happen as shown in \textbf{Figs. \ref{cobra} (c)}. More details can be found in the Supplementary Material. As is well known, spin transport can be mediated not only by conduction electrons, but also by magnons in ferromagnetic materials \cite{30, 31, 32}. Thus, the local magnetic anisotropies contribute to the electron-magnon interactions in ferromagnetic bilayers. The electron-magnon interactions induce additional spin-flip processes, as discussed by Y. Omori et. al \cite{33}. Such spin-flip processes are equivalent in magnitude for up-to-down and down-to-up spin channels in ferromagnetic systems \cite{34}. In our findings, such asymmetric scatterings which are spin-dependent would contribute to the anti-polarization of the spin current. In other words, this effect is associated with the distribution of small anisotropies capable of blocking the inversion of the spin current polarization, producing an anti-polarization of spin current due to the coupling between local magnetic anisotropies of the Py/Co bilayers. The relationship between charge and spin currents is well known $\stackrel{\leftrightarrow}{J}_{C} = (2e/\hbar)\theta_{SH} \stackrel{\leftrightarrow}{J}_{S} \times \hat{\sigma}$ [4, 5, 6, 7, 8, 9, 10, 11, 12, 13, 14]. For positive dc current our associated charge current is
\begin{equation} 
	\stackrel{\leftrightarrow}{J}_{C}^{(+)} = \left(\frac{2e}{\hbar}\right)\theta_{SH} \stackrel{\leftrightarrow}{J}_{S} \times \hat{\sigma},
	\label{1}
\end{equation}
and for the negative dc current our charge current is 
\begin{equation} 
	\stackrel{\leftrightarrow}{J}_{C}^{(-)} = \left(\frac{2e}{\hbar}\right)\theta_{SH} \stackrel{\leftrightarrow}{J}_{S} \times \hat{\sigma}.
	\label{2}
\end{equation}
The equations (1) and (2) reveal the following property $(\stackrel{\leftrightarrow}{J}_{C}^{(+)})^{\dagger} = - \stackrel{\leftrightarrow}{J}_{C}^{(-)}$. Thus, by the hermiticity property $(\stackrel{\leftrightarrow}{J}_{S})^{\dagger} = \stackrel{\leftrightarrow}{J}_{S}$, which shows the hermiticity property of the spin current generated by its anti-polarization. Here, the tensor current represents the current in general and not a component.

In summary, a phenomenon is presented here, which we call anti-polarization of spin current due to the coupling between interfacial magnetic anisotropies of Py/Co bilayers, where Co films were deposited at a temperature of 300 °C. This effect allows the observation of spin current hermiticity. It is unequivocally shown that the spin current changes its polarization when it is being excited via negative dc current, such blocking was associated with the anisotropic distribution of the two materials at the interface. We believe that such a phenomenon will be relevant for the manipulation of information in spintronic devices.

\section*{Data Availability Statements}

The data that support the findings of this study are available from the corresponding author upon reasonable request.

\section*{Supplementary Material}

Ferromagnetic resonance signals were obtained using the VNA for one magnetic field applied from H = 0.5 kOe, resonance frequency from 8.7 GHz, and dc currents from I = +1.8, +0.8, 0.0, -0.8 and -1.8 mA.

\section*{Acknowledgements}

This research was supported by Conselho Nacional de Desenvolvimento Científico e Tecnológico (CNPq), Coordenação de Aperfeiçoamento de Pessoal de Nível Superior (CAPES), Financiadora de Estudos e Projetos (FINEP), and Fundação de Amparo à Ciência e Tecnologia do Estado de Pernambuco (FACEPE).

\bibliographystyle{MiKTeX}

\end{document}